\documentclass[aps,prb,twocolumn]{revtex4}
\usepackage{graphicx}

\begin{document}
\title{Theoretical study of doped-Tl$_{2}$Mn$_{2}$O$_{7}$ and Tl$_{2}$Mn$_{2}$O$_{7}$ under pressure}
\author{Molly De Raychaudhury$^{1}$, T. Saha-Dasgupta$^{1}$ and D. D. Sarma$^{2,3,}$$^*$}
\affiliation{$^1$ S.N. Bose National Centre for Basic Sciences,
Kolkata 700098, India} \affiliation{$^2$ Indian Association for the
Cultivation of Science, Jadavpur, Kolkata 700032,
India}\affiliation{$^3$ Solid State and Structural Chemistry Unit,
Indian Institute of Science, Bangalore-560012, India}
\date{\today}

\begin{abstract}
Using first-principles density functional based calculations, we
study the effect of doping and pressure in manganese based
pyrochlore compound,Tl$_{2}$Mn$_{2}$O$_{7}$ that exhibits colossal
magneto-resistive behavior.  The theoretical study is motivated by
the counter-intuitive experimental observation of suppression of
ferromagnetic transition temperature upon
application of pressure and its enhancement upon substitution of Mn by
moderate amount of nonmagnetic Sb ion. We also attempt to resolve the issue
related to crystal structure changes that may occur upon application
of pressure.
\end{abstract}
\maketitle
\noindent
\section{Introduction}
Manganese-based pyrochlore compounds such as Tl$_{2}$Mn$_{2}$O$_{7}$, form an interesting class of compounds
which despite having similar colossal magneto-resistive (CMR) effect
exhibit markedly different features compared to that of manganites.
In contrast to manganites, these compounds are not mixed-valent, do not have any appreciable Jahn-Teller distortion \cite{hwang} and seemingly show metallic behavior
\cite{shima} both below and above the magnetic transition
temperature T$_c$ ($\approx$ 140 K). Concerning the metallic behavior above T$_{c}$ though, it is not totally clear 
whether it is a good metal or more like a bad metal or an insulator-like compound which arises due to
difficulties related to sample preparation and defects present in the paramagnetic phase. Though the original
data by Shimakawa et. al. \cite{shima} suggests the metallic behavior above T$_c$, the optical conductivity 
data\cite{note-insulator1} and and photoemission and x-ray absorption\cite{note-insulator2} near edge 
spectroscopy point towards an insulator-like behavior in the paramagnetic regime. These
non-manganite like features rule out the generally applicable double
exchange mechanism or polaronic mechanism as the driving mechanism
of ferromagnetism in this class of compound, and remained a mystery
for quite some time. In absence of double exchange mechanism, a
strong contestant for the driving mechanism is the Mn-O-Mn mediated
superexchange mechanism, particularly since Mn-O-Mn form an angle
\cite{struc} of 133$^{o}$, significantly smaller than 180$^{o}$.
This is however contradicted by the experimental findings that the
ferromagnetic T$_c$ gets suppressed by application of pressure
\cite{sushko} and gets enhanced by moderate amount of introduction
of nonmagnetic ion like Sb in place of Mn \cite{alonso}. Employing
N-th order muffin tin orbital (MTO) based NMTO-downfolding
technique\cite{nmto}, we have recently established~\cite{dasgupta}
that driving mechanism of ferromagnetism in this class of compounds
is neither super-exchange or double-exchange, nor any exotic
mechanism with nearest neighbor (NN) antiferromagnetic (AFM)
coupling dominated by long-ranged ferromagnetic (FM)
interaction\cite{super_new}, but a kinetic energy assisted mechanism
similar to that of double perovskite compounds like Sr$_{2}$FeMoO$_6$, Sr$_{2}$FeReO$_6$\cite{prl-srfemo,dd} and Mn-doped GaAs\cite{priya}.

\par
In the present paper, we give a detail account of the doping
effect and the pressure effect from the electronic structure point
of view, which lend further support to the existence of the
proposed kinetic energy driven mechanism in these compounds. While
the changes that happen in the crystal structure upon doping is
clear, the situation is rather ambiguous in case of application of
pressure. The initial measurement hinted to a decrease in Mn-O-Mn
bond angle \cite{sushko,shimakawa_priv} alongwith the compression
of Mn-O bondlength. However the subsequent measurement
\cite{valesco67} indicated an increase in Mn-O-Mn bond angle which
was used in terms of super exchange mechanism to explain the
decrease in T$_c$. In view of these conflicting claims, we also
take the opportunity to theoretically address the effect of
pressure on the structure of Tl$_{2}$Mn$_{2}$O$_{7}$ and its
implication on magnetism by detailed electronic structure
calculations together with structural relaxations under pressure.

The rest of the paper is organized in the following way. Section II deals with various methodologies that have
been employed to compute the crystallographic structure (in case of Tl$_2$Mn$_2$O$_7$ under pressure) and to
compute and analyze the subsequent electronic structure of Tl$_2$Mn$_2$O$_7$ under doping and pressure. In section
III, we present our results. This section is divided into several subsections. In section IIIA, the electronic
structure of pristine Tl$_2$Mn$_2$O$_7$ is presented and the proposed kinetic energy driven mechanism is introduced
as a reference, which will form the basis of discussion in the following. Section IIIB is devoted to  Tl$_2$Mn$_2$O$_7$ under pressure ,while section IIIC is devoted to Sb-doped Tl$_2$Mn$_2$O$_7$. Finally, we conclude
the paper with a summary presented in section IV.

\section{Methodology}

We have investigated the electronic structure of pure Tl$_{2}$Mn$_{2}$O$_{7}$, Tl$_{2}$Mn$_{2-x}$Sb$_{x}$O$_{7}$
and Tl$_{2}$Mn$_{2}$O$_{7}$ under pressure by the tight-binding linear muffin-tin orbital (LMTO)\cite{tb-lmto}
method, computed within the framework of the local spin density approximation (LSDA) of the density functional
theory (DFT). The employed basis set, consisted of Tl-$6s$, $6p$ and $5d$ states, Mn-$4s$, $4p$ and $3d$ states,
and O-$2s$ and $2p$ states. Three different classes of empty spheres were used to fill up the space.
The self-consistent calculations were performed on a BZ mesh of 12$\times$12$\times$12. 
\par
The computed
band structures were analyzed and interpreted in terms of NMTO-downfolding technique \cite{nmto}.
The NMTO technique, introduced and implemented in recent years, goes beyond the scope of standard LMTO technique
by defining an energetically accurate basis with consistent description throughout the space involving both the MT
spheres and the interstitial. As important feature of this technique is the energy-selective downfolding procedure
where one integrates out degrees of freedom and keeps only few degrees of freedom active to define a few-orbital,
low-energy Hamiltonian. The effective NMTOs defining such Hamiltonian, serve as the Wannier or
Wannier-like function corresponding to the selected low-energy bands.

For the calculation of the exchange interaction strengths, J, a supercell involving a sixteen Mn atoms was
constructed which amounts to enlarging the original unit cell by four times. This was achieved by replacing
the original face centered cubic (FCC) lattice by the simple cubic lattice with four times as many atoms as that
in FCC. The basic structural unit of Mn sublattice is the Mn$_4$ tetrahedra which share corners to form an
infinite 3D lattice. We calculated the LSDA total energies by considering different magnetic configurations
of this sublattice. In total, six spin configurations have been chosen. One of the six configurations was FM and
the other five were AFM. The details of the spin configurations chosen are as given in Table I in ref\cite{dasgupta}.
The total energies are then mapped on to an effective Heisenberg Hamiltonian formed by the Mn$^{+4}$ spins.
The Hamiltonian considered till the 3NN(third nearest neighbor) interactions is:
\begin{equation}
H=J_{1}\sum_{nn}S_{i}.S_{j}+J_{2}\sum_{2nn}S_{i}.S_{j}+J_{3}\sum_{3nn}S_{i}.S_{j}
\end{equation}
where S$_{i}$ denoted the spin-3/2 operator corresponding to the
Mn $^{+4}$ spins at the site i, and {\it J$_{1}$,J$_{2}$} and {\it
J$_{3}$} denote the NN, 2NN(second nearest neighbor) and 3NN magnetic exchange interaction
strengths. Since we have restricted ourselves to only five AFM
configurations amongst many possible AFM configurations, we
computed seven different estimates of these J's along with their
standard deviations employing set of three energy differences
chosen out of total five different energy differences. This
limitation can be overcome by taking as many AFM spin
configurations as practically possible. The mean-field estimate of
T$_{c}$ is given by
\begin{equation}
 T^{mf}_{c}={\frac{S(S+1)}{3k_{B}}}J_{0}
\end{equation}
 where {\it J$_{0}$}, is the  net effective magnetic exchange interaction, given by
\begin{equation}
J_{0}=z_1 J_{1}+ z_2 J_{2}+ z_3 J_{3}
\end{equation}
and $S={\frac{3}{2}}$. k$_{B}$ is the Boltzman constant and $z_1, z_2, z_3$ are the number of NN, 2NN and 3NN
Mn pairs. For undoped Tl$_{2}$Mn$_{2}$O$_{7}$, $z_1$ = 6, $z_2$ = 12 and $z_3$ = 12. However, these numbers
change in case of doped calculations where some of the Mn atoms gets replaced by the nonmagnetic
Sb ion and therefore do not contribute in the calculation of magnetic energy. The application of mean-field theory is naturally expected to overestimate the T$_{c}$. Use of Monte Carlo simulation \cite{mc} or Green Function method \cite{gf} can give better estimates and agreements with experimentally measured values. One needs to keep this in mind when comparisons are being made with the measured T$_{c}$ in the following sections.
\begin{figure*}
\includegraphics[width=16.0cm,keepaspectratio]{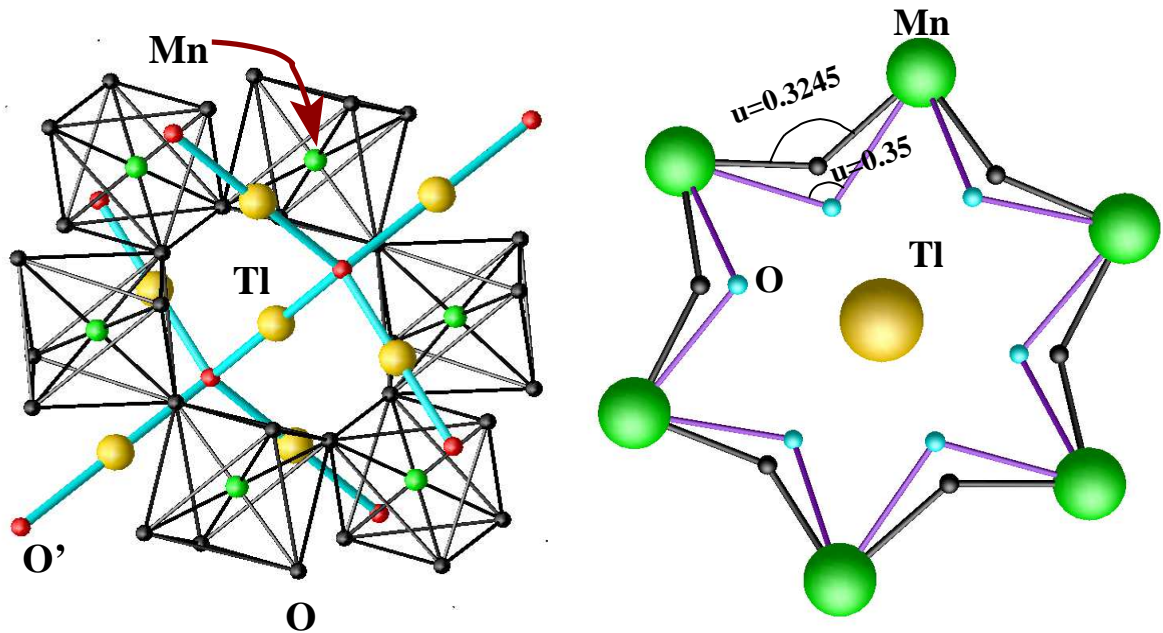}
\caption{(color on-line) Left panel: The Tl$_{2}$Mn$_{2}$O$_{7}$ structure consisting of the ring formed by MnO$_6$ octahedra with Tl-O' rods passing through the holes of the rings. Right panel: Backbone of the structure realized by the ring like structure formed by the Mn-O-Mn network (the Tl-O' rod network is not shown for clarity). Gold, green and black or cyan spheres denote the Tl, Mn and O atoms. Black and cyan spheres correspond to oxygen (O) atoms with two different oxygen (O) position parameter {\it u}.}
\label{structure}
\end{figure*}

Tl$_2$Mn$_2$O$_7$ occurs in cubic, face-centered lattice with $Fd-3m$ space group and two formula unit in the cell.\cite{struc}
There are two types of oxygen, O' and O. While O oxygen provide the octahedral surrounding of the Mn atom
which corner share to give rise to a geometry of a closed ring or cage, O' atoms form nearest neighbor bonds with
Tl giving rise to Tl$_{2}$O' complex running through the cage like geometry as shown in left panel of Fig. 1.
The structural parameters for the pristine Tl$_{2}$Mn$_{2}$O$_{7}$, and Sb-doped Tl$_{2}$Mn$_{2}$O$_{7}$ have been
taken as the experimentally determined structure published in literature \cite{magmom,alonso}. However, as
mentioned already, the structural information for Tl$_{2}$Mn$_{2}$O$_{7}$ under pressure is somewhat controversial
which calls for structural optimization. For this purpose, we have adopted the pseudopotential method\cite{vanderbilt} as
implemented
in the VASP\cite{vasp} code. We have employed the  pseudopotentials generated by the projector-augmented wave (PAW)\cite{paw} method. Wavefunctions have been
expanded in terms of plane waves with a cut off energy of 400 eV. A {\it k}-mesh of 6$\times$6$\times$6 was employed to perform
the self-consistent calculations. In Tl$_{2}$Mn$_{2}$O$_{7}$, Tl, Mn, O' and O atoms occupy the Wyckoff positions,
16d, 16c, 8b and 48f respectively. Among these only the positions given by 48f contain the internal parameter,{\it u}. This parameter {\it u} defines the $\angle$Mn-O-Mn (see right panel of Fig.\ref{structure}). As {\it u} increases,
the $\angle$Mn-O-Mn decreases and vice versa. When pressure is applied, it needs to be investigated whether the
{\it u} parameter also varies along with the reduction in the bondlengths. If at all it changes, it has to be
ascertained in what way it does so. In our calculation, the O position is relaxed by varying {\it u}, in order
to find out the variation of total energy with angle formed between Mn, O and Mn atoms. In order to obtain the
theoretical ground state, cohesive properties and estimate the pressure, Murnaghan\cite{murna} equation of state has
been employed to fit the LSDA total energies. The Murnaghan equation of state is given by
\begin{equation}
E(V)=E_{0}+\frac{B_{0}.V}{B_{0}^{'}}\left[\frac{(V_{0}/V)^{B_{0}^{'}}}{B_{0}^{'}-1}+1\right]-
\frac{B_{0}.V_{0}}{B_{0}^{'}-1}
\end{equation}
where {\it V$_{0}$} is the equilibrium volume, {\it B$_{0}$} is
the bulk modulus and is given by $B_{0}=-V(\delta P/\delta V)_{T}$
evaluated at volume {\it V$_{0}$}. {\it B$^{'}_{0}$} is the
pressure derivative of {\it B$_{0}$} also evaluated at volume {\it
V$_{0}$}. {\it B$^{'}_{0}$} provides a measure of stiffness of the
material upon increasing pressure. Typical values for the
parameter {\it B$^{'}_{0}$} are between 4 and 7. Pressure, {\it P},
corresponding to volume, {\it V}, occupied by a cell of a
particular lattice parameter is given by
\begin{equation}
P(V)={\frac{B_{0}}{B_{0}^{'}}}\left[{{(V_{0}/V)}^{B_{0}^{'}}}-1\right]
\end{equation}

\section{Results}

\subsection{FM Electronic Structure of Pristine Tl$_2$Mn$_2$O$_7$ and the Underlying Mechanism}
\begin{figure}[ht]
\includegraphics[width=8.5cm,keepaspectratio]{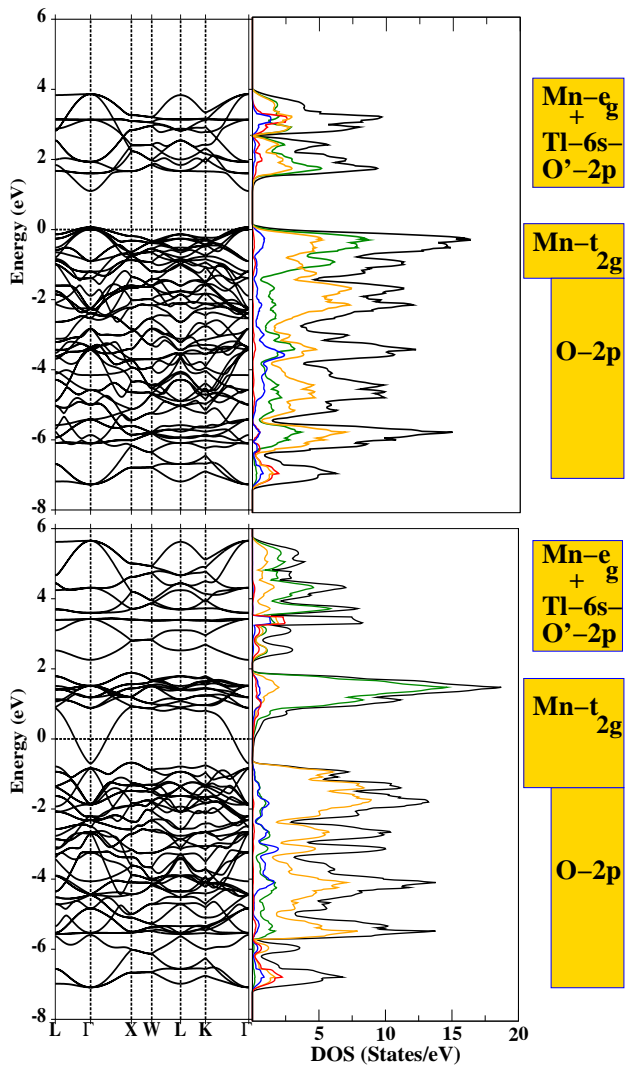}
\caption{(color on-line) Left panels show the spin-polarized band
structure computed within LSDA for Tl$_2$Mn$_2$O$_7$ . The zero of
the energy is set at the LSDA Fermi energy, E$_F$. Right panels
give the spin-polarized total (black), Tl-6s (red), Mn-3d
(green), O-2p (orange) and O'-2p (blue) density of states. Upper
and the lower figures correspond to the majority and minority spin
contributions respectively in both the panels. Bars 
on the right signify the energy range for which various characters dominate.}
\label{bands+dos}
\end{figure}

The LSDA electronic structure of ferromagnetically ordered
Tl$_2$Mn$_2$O$_7$, computed with TB-LMTO method is shown in
Fig.\ref{bands+dos}. This figure shows both the band structure and
the orbital projected density of states (DOS). Consistent with the
nominal valence of O$^{2-}$ and Mn$^{4+}$, oxygen dominated states
are totally occupied, while the crystal field split Mn $t_{2g}$
states are occupied in the majority spin channel and close to empty
in the minority spin channel. Mn $e_g$ states consistent with $3d^3$
configuration of Mn, remain empty in both the spin channels. Tl
bands, consistent with Tl$^{3+}$ nominal valence also remain almost
empty. However,  interestingly in the majority spin channel, the
Tl-6s state mixed with O'-2p states overlap with Mn $e_g$ manifold
spanning an energy range of about 2-4 eV\cite{hybrid}, while in the minority spin
channel, they overlap with nearly empty Mn $t_{2g}$ dominated
states, which give rise to highly dispersive band in bottom of this
manifold, crossing the Fermi level, E$_F$, as seen in lower, left
panel of Fig.\ref{bands+dos}. This hints to a large spin-splitting
at the Tl site. This is unusual in the sense, one would normally
expect Tl to be essentially nonmagnetic with tiny spin splitting.
This unusual feature plays the key role in providing the
understanding of the driving mechanism of ferromagnetism in
Tl$_2$Mn$_2$O$_7$. This has been explained in detail in
ref\cite{dasgupta}. As has been shown in ref\cite{dasgupta}, the
unhybridized effective Tl-O' level is positioned in between the
spin-split Mn $t_{2g}$ states and on turning on the hybridization,
it induces a renormalized spin-splitting within the Tl-O' level
which is directed in a opposite way compared to the spin-splitting
at the Mn site. This happens due to pushing up of the Tl-O' spin-up
state and pushing down of the Tl-O' spin-down states, driven by
coupling with the Mn $t_{2g}$ states of the same symmetry. The
energy gain contributed by this hybridization induced negative spin
polarization of the otherwise nonmagnetic element, is the central
concept in this novel mechanism. This mechanism ensures a specific
spin orientation between the localized Mn $t_{2g}$ spins and the
mobile carrier formed by the hybridized renormalized state, which in
turn aligns the spins in Mn sublattice. This is a general mechanism
and found to be operative in case of a number of compounds like
Sr$_{2}$FeMoO$_6$, Sr$_{2}$FeReO$_6$\cite{prl-srfemo,dd}, Mn-doped GaAs\cite{priya} and
CuCr$_2$S$_4$\cite{de2} etc.

In the following subsections, we will present a detail account of the changes in the electronic structure  and the variation in T$_c$ of
Tl$_2$Mn$_2$O$_7$ with doping and under pressure. We will argue that the
counter-intuitive variation in T$_c$ has its origin in the modified electronic structure and
follows naturally from the framework of proposed hybridization induced mechanism of ferromagnetism as
was established in ref\cite{dasgupta} and reviewed in the above.

\subsection{Tl$_2$Mn$_2$O$_7$ under pressure}

As explained before, to investigate Tl$_2$Mn$_2$O$_7$ under pressure one needs to first settle the issues related
to crystal structure. To be precise, does applied pressure increase or decrease or not alter the $\angle$Mn-O-Mn ?
For this purpose, we have carried out a series of structural optimization calculations with pseudopotential method \cite{vanderbilt}. At first the cohesive properties were calculated within LSDA. Following the recent claim\cite{rabe} that
application of on-site U can also influence the cohesive properties of transition metal oxides, we have
repeated our calculations also with LSDA+U functional.

\begin{figure}[th]
\includegraphics[width=8.5cm,keepaspectratio]{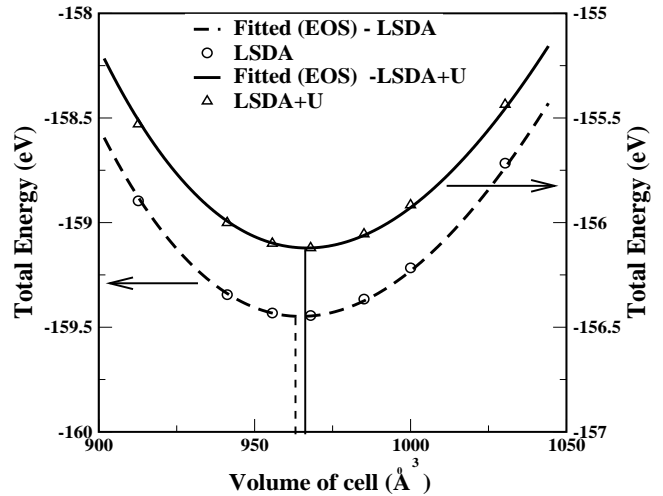}
\caption{Total energy vs. volume of the unitcell calculations
within LSDA and LSDA+U. The symbols denote the DFT total energies
and the solid and dashed lines depict the total energies obtained
from fitting with Murnaghan \cite{murna} equation of state
(EOS).}
\label{bulk}
\end{figure}
LSDA total energy vs volume calculation, shown in Fig.\ref{bulk}, has a minimum corresponding to
lattice constant of 9.881 $\AA$.  Considering the experimental value of the lattice constant is 9.892 $\AA$,
this gives a deviation of about 0.1 $\%$, which is within the limit of LSDA overbinding. The value of
the bulk modulus, obtained by fitting the total energy vs volume curve with Murnagahan
equation of state \cite{murna} is 230.6 GPa, which is of the same order as the experimentally
measured value of $\approx$200 GPa in Tl$_{1.8}$Cd$_{0.2}$Mn$_{2}$O$_{7}$\cite{valesco67}. B$^{'}_{0}$ is +4.84, much higher than that in  Cd doped Tl$_{2}$Mn$_{2}$O$_{7}$ but within the expected limit. Applying an on-site Coulomb interaction
of U=2 eV and J=0.7 eV on the Mn site leads to a slightly better agreement with the experimental values. The lattice parameter
changes to 9.888 $\AA$, B$_{0}$ and B$^{'}_{0}$ remain unaltered. In view of the slightly improved results of LSDA+U,
we have used the B$_0$ and  B$^{'}_{0}$ as estimated by LSDA+U calculations to compute the pressure as
given by Eq.(5).

\begin{table}[htbp]
\begin{tabular}{|cccc|} \hline
& LDA & LSDA+U & EXPT.\\ \hline
a($\AA$)& 9.881 & 9.888 & 9.892 \\
{\it u} & 0.3214 & 0.3227 & 0.3254\\
Mn-O ($\AA$)& 1.886 & 1.890 & 1.901\\
$\angle Mn-O-Mn$(deg) & 136.01 & 135.29 & 133.8 \\
B$_{0}$ (GPa) & 230.6 & 230.7 &\\
B$^{'}_{0}$ & +4.84 & +4.85 &\\
\hline
\end{tabular}
\caption{Calculated ground state structural parameters of Tl$_{2}$Mn$_{2}$O$_{7}$ within LSDA and LSDA+U compared with the experimental\cite{struc} data at ambient condition.}
\label{AMBIENT}
\end{table}

\par
\begin{figure}[ht]
\includegraphics[width=8.5cm,keepaspectratio]{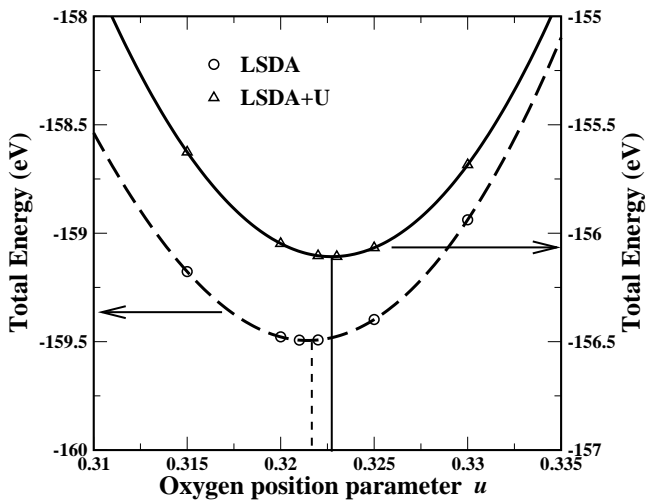}
\caption{Relaxation of the Oxygen-O position parameter {\it u} at ambient condition within LSDA and LSDA+U in Tl$_{2}$Mn$_{2}$O$_{7}$. The symbols are the DFT total energies and the solid and dashed lines are cubic spline interpolated.}
\label{bulku}
\end{figure}

Till now we have considered the experimental $\angle$Mn-O-Mn as the one for the theoretical ground state. One needs to
relax this angle in order to obtain the true theoretically predicted ground state. The co-ordinate of O atom in the basis is given
by ({\it u},.125,.125). The {\it u} parameter will push O atom in or out depending on its value, thereby
changing the angle formed between Mn, O and Mn atoms. The {\it u} parameter is relaxed in the subsequent
calculations, and the LSDA (for 9.881 $\AA$) and LSDA+U total energies (for 9.888 $\AA$) are computed varying
the {\it u} parameter. The minimum in the E vs {\it u} curve (see Fig.\ref{bulku}) is obtained for u=0.3214
and 0.3227 respectively. The $\angle$Mn-O-Mn in the ground state obtained within LSDA and LSDA+U framework are
given in Table I  along with the corresponding experimental values. Slightly better results are obtained for the
LSDA+U calculations, a trend already observed in  calculation for equilibrium lattice constant. The result match
with experimental values upto the second decimal point for {\it u} and within 1.5$^{o}$ for Mn-O-Mn angle.
\begin{figure}[ht]
\includegraphics[width=8.5cm,keepaspectratio]{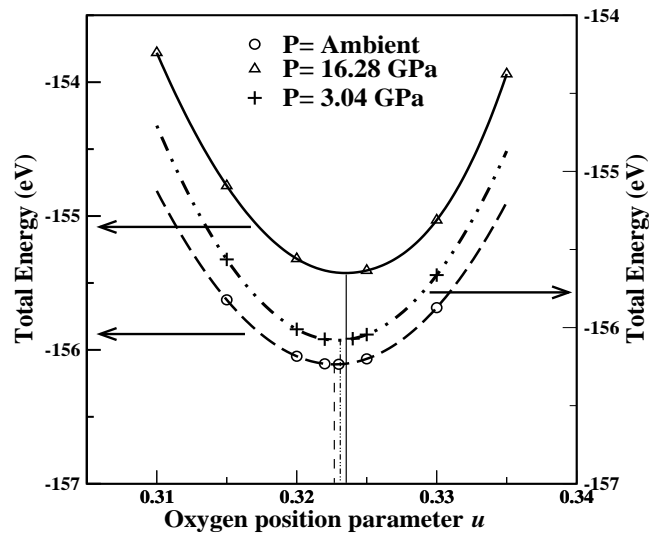}
\caption{Total energy vs {\it u}, i. e. relaxation of the Oxygen (O) position parameter {\it u} at ambient condition and under pressure of 3.04 and 16.28 GPa within LSDA+U in Tl$_{2}$Mn$_{2}$O$_{7}$. The symbols are the DFT total energies and the solid, dashed and dashed-dotted lines are cubic spline interpolated.}
\label{relaxfig}
\end{figure}

\par
Having established the ground state, in the next step we relax the
O position for lattice parameters a=9.846 and 9.69 $\AA$ (LSDA+U)
which amount to applying a pressure of 3.04 GPa and 16.28 GPa.
This will clarify whether the Mn-O-Mn angle also changes with the
decrease in the interatomic distances. The minimum obtained from
the total energy vs. {\it u} plots, a shown in Fig.
\ref{relaxfig} are listed in Table II. We find that on
application of pressure, the $\angle$Mn-O-Mn reduces
$insignificantly$, by about 0.1$^{o}$ and 0.4$^{o}$ for pressure
of 3.04 GPa and 16.28 GPa respectively. A reduction of the angle
by about 0.6$^{0}$ occurs for the same applied pressure (16.28
GPa) within LSDA. We thereby establish that, regardless of kind of
DFT one is using, on applying pressure $\angle$Mn-O-Mn does not
increase rather decreases though almost insignificantly. The major
effects comes from reduction in the bondlengths. In the following calculations, we have therefore kept the $\angle$Mn-O-Mn unchanged upon application of pressure.
\begin{table}
\begin{tabular}{|cccc|} \hline
& AMBIENT & 3.04 GPa & 16.28 GPa\\ \hline
{\it u} & 0.3227 & 0.3230 & 0.3234 \\
Mn-O($\AA$) & 1.890 & 1.883 & 1.854\\
$\angle Mn-O-Mn$ (deg)& 135.29 & 135.18 & 134.90 \\
\hline
\end{tabular}
\caption{Structural parameters obtained within LSDA+U at ambient and two different pressures upon relaxation of the Oxygen(O) position parameter {\it u} in Tl$_{2}$Mn$_{2}$O$_{7}$.}
\label{Relaxation}
\end{table}
\par
We have carried out LSDA total energy calculations for various spin arrangements of Mn atoms as explained
in section II. The relative LSDA total energies, as given in Table III are mapped on to Eq. (1) in order to extract the J's.
\begin{table}[htbp]
\begin{tabular}{|c|ccc|} \hline
& &$\Delta$E(meV)& \\ \hline
& Normal & 3.04 GPa & 16.28 GPa\\ \hline
FM & 0 & 0 & 0\\
AFM1 & 9.55 & 9.18  & 7.36\\
AFM2 & 13.98 & 14.17 & 13.05\\
AFM3 & 12.24 & 11.49 & 8.82\\
AFM4 & 17.59 & 22.5 & 16.24\\
AFM5 & 18.39 & 17.7 & 14.04\\
\hline
\end{tabular}
\caption{Relative LSDA energies per Mn ion in meV in FM and five AFM spin configurations for Tl$_{2}$Mn$_{2}$O$_{7}$ at normal, 3.04 GPa and 16.28 GPa pressures. All energies are converged upto 0.01 meV per Mn ion.}
\end{table}
The magnetic exchange interaction strengths are calculated and shown in Table IV for almost 2 $\%$ reduction in experimental lattice parameter, amounting to a pressure of 16.28 GPa. Considering pressure of 16.28 GPa, we observe a substantial decrease in the NN FM interaction. The 2 and 3 NN interactions remain almost the same. This leads to a mean-field T$_{c}$ of about 140 K, leading to a decrease of about 41 K. The gradient is about -2.5 K/GPa.
The gradient calculated from another set of calculations at 3.04 GPa is about -2.3 K/GPa. One can approximate a linear relation between T$_{c}$ and applied pressure. Thus the average gradient is about -2.4 K/GPa. This is not an unreasonable estimate considering the mean-field overestimation. The ratio of exact T$_{c}$ and $T^{mf}_{c}$ is estimated as 0.79 and 0.81 for BCC and FCC lattice with coordination 8 and 12 respectively. Considering the average coordination of 10 for Mn sublattice in Tl$_{2}$Mn$_{2}$O$_{7}$, one would expect the gradient in T$_c$ to be about -1.9 K/GPa. This is a bit  larger than the experimental values, which are -1.6 K/GPa\cite{sushko} and about -1 K/GPa \cite{valesco67} . A plausible explanation for this somewhat overestimation may be due to computational limitations imposed by the number of spin configurations. A better estimate of the gradient might have been reached had all the possible spin configurations been considered for calculation of the J's. Therefore one can conclude that within the limitation of our method, we have established the trend in ferromagnetism of Tl$_{2}$Mn$_{2}$O$_{7}$ with applied pressure correctly.

\begin{table}[htbp]
\begin{tabular}{|cccccc|} \hline
Pressure Appl.& J$_{1}$ & J$_{2}$ & J$_{3}$ & J$_{0}$ & T$^{mf}_{c}$\\ \hline
(GPa)& (meV) & (meV) & (meV) & (meV) & (K) \\ \hline
0& -2.52 & -0.11 & +0.33 & 12.48 & 181\\
3.04 & -2.48 & -0.16 & +0.40 & 12.06 &174\\
16.28 &-1.99 &-0.16 &+0.35 & 9.66 & 140 \\
\hline
\end{tabular}
\caption{NN, 2NN, 3NN and effective exchange interaction strengths and mean-field Curie Temperature calculated within LSDA at ambient condition and at 3.04 and 16.28 GPa for Tl$_{2}$Mn$_{2}$O$_{7}$.}
\label{J_pressure}
\end{table}
\begin{figure}[th]
\includegraphics[width=8.5cm,keepaspectratio]{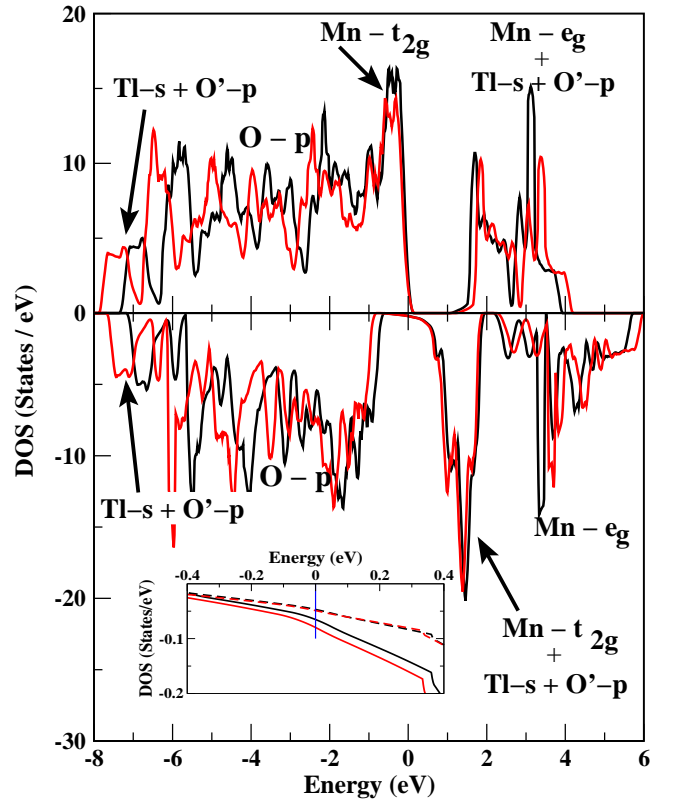}
\caption{(color on-line) Spin-polarized density of states at
ambient condition (black) and 16.28 GPa (red) calculated within
LSDA for Tl$_{2}$Mn$_{2}$O$_{7}$.  The upper (lower) panel corresponds to the majority (minority) spin. The negative of density of states has been plotted in the lower panel for clarity. Inset shows minority spin
channel density of states for the energy region near Fermi level
at ambient and 16.28 GPa. The solid and dashed lines show the
contribution from Mn - 3d and Tl-6s+O'-2p states respectively
within LSDA.}
\label{pressureplot}
\end{figure}

In Fig.\ref{pressureplot} we show the density of states of
Tl$_2$Mn$_2$O$_7$ under 16.28 GPa pressure alongwith that of
Tl$_2$Mn$_2$O$_7$ under ambient pressure for comparison. Upon
application of pressure, the bondlengths shorten which enhances the
Mn-O and Tl-O' interaction and as a result net bandwidth expands and 
t$_{2g}$-e$_g$ splitting enhances. However the key effect is seen in
the inset, where we show the Mn-d and Tl-s projected density of
states in an energy window close to E$_F$ in solid and dashed lines
respectively. They are shown in two different colors, corresponding
to ambient and 16.28 GPa pressure. We notice that the relative
proportion of Tl-O' in the hybridized, dispersive band has reduced
significantly upon application of pressure. It has reduced from 45
$\%$ in case of ambient pressure to 36 $\%$ in case of 16.28 GPa
pressure. This indicates reduction of Mn-Tl-O' hybridization upon
application of pressure. Since the underlying mechanism of
ferromagnetism is based on the Mn-Tl-O' hybridization, weakening of
such hybridization would cause a decrease in T$_c$ as supported by
the detail calculation presented above. The change in the
hybridization is proportional to the change in the hopping
interaction strength, $t$ and inversely to the change in the energy
level separation $\Delta$. Application of pressure causes shortening
of bondlength, hence one would expect hopping to increase and in
turn the hybridization to increase, opposite to what is observed in
the calculation. For this we performed the NMTO-downfolding
calculation keeping Mn-3d, Tl-6s and O-2p degrees of freedom active
and integrating out the rest, including O' degrees of freedom to
define a Tl-O' effective level. The real-space representation of
that downfolded Hamiltonian in the NMTO-Wannier function basis for
Tl$_2$Mn$_2$O$_7$ under application of 16.28 GPa pressure shows the
Tl-O' effective level to shift down by 0.3 eV compared to that under
ambient pressure. This increases the $\Delta$ and substantially
reduces the Tl-O'-Mn hybridization, substantial enough to compensate
the increase in t$_{sd\sigma}$ hopping interaction between
Mn-t$_{2g}$ and the effective Tl-$s$ orbital and to produce a net
negative contribution to the hybridization.This is an interesting, counter-intuitive
situation which can only be unraveled by a detail analysis of the underlying electronic structure as has been done in the present study.

\subsection{Sb-doped Tl$_2$Mn$_2$O$_7$}

To mimic the doping effect, we have carried out calculations using supercell containing sixteen Mn atoms. Construction
of such supercell has been described in section II. Out of these, sixteen Mn atoms, some are substituted by Sb
to simulate the effect of doping. There are two doping levels that have been investigated. In one case,
one of the Mn atom is substituted by an Sb atom, and in another case, two of the Mn atoms are replaced by
two Sb atoms. The former amounts to doping level of x=0.125 or 6.25$\%$ and the latter amounts to
doping of x=0.25 or 12.5 $\%$. These are not exactly the doping levels that has been investigated experimentally
but close to them ({\it i.e.} x = 0.1 and 0.2 respectively).

Substitution of Mn$^{4+}$ by larger cation Sb$^{5+}$ does not cause any significant changes in the crystal
structure, other than the fact that lattice expands slightly keeping the various angles unchanged. The lattice parameters
corresponding to doping levels of x=0.125 and 0.25 are obtained by interpolation of the experimentally determined
lattice constants at various doping as shown in the inset of Fig.1. in ref\cite{alonso}. The lattice parameters
for x =0.125 and 0.25 are increased by 0.3$\%$ and 0.9$\%$ respectively, compared to undoped lattice constant.

To estimate the exchange interaction strengths for  Tl$_{2}$Mn$_{2-x}$Sb$_{x}$O$_{7}$ with x=0.125 and 0.25
we have carried out LSDA total energy calculations for various spin arrangements of Mn atoms as explained
in section II. The relative LSDA total energies, as given in Table V are mapped on to Eq. (1) in order to extract the J's. 
\begin{table}[htbp]
\begin{tabular}{|c|ccc|} \hline
& &$\Delta$E (meV)& \\ \hline
& x=0 & x=0.125 & x=0.25\\ \hline
FM & 0 & 0 & 0\\
AFM1 & 9.55 & 16.50 & 23.96\\
AFM2 & 13.98 & 26.07 & 29.39\\
AFM3 & 12.24 & 23.18 & 31.53\\
AFM4 & 17.59 & 33.30 & 35.69\\
AFM5 & 18.39 & 34.39 & 40.11\\
\hline
\end{tabular}
\caption{Relative LSDA energies per Mn ion in meV in FM and five AFM spin configurations for doped Tl$_{2}$Mn$_{2}$O$_{7}$. All energies are converged upto 0.01 meV per Mn ion.}
\end{table}
At 6.25 $\%$ doping the average coordination numbers change to 5.6, 11.2 and 11.2 respectively.  In order to check
whether the chosen disordered configuration has any influence on the trend, we have repeated the calculation
with two Sb atoms in the unit cell for two cases. In one case, two Sb atoms were chosen as NN pairs while
in the second case they have been chosen as 3NN pairs. The respective coordination numbers at 12.5 $\%$ doping are 5.14, 10.28 and 10.86 for 3NN substitution and 5.28, 10.28 and 10.28 for NN substitution. In all the three cases, i.e for x=0.125, 0.25 (NN) and
0.25 (3NN), the ground state is found to be FM.

\begin{table}[htbp]
\begin{tabular}{|cccccccc|} \hline
&doping & J$_{1}$ & J$_{2}$ & J$_{3}$&J$_{0}$& D(E$_{F}$)& $T^{mf}_{c}$\\ \hline
&conc. & (meV) & (meV) & (meV) & (meV) & (States/eV) & (K)\\ \hline
&x=0.0 & -2.52 & -0.11 & +0.33 & -12.48 & 0.2 &181\\
&x=0.125 & -3.65& -0.12 & -0.41 & -26.38 & 1.0 & 382\\
&x=0.25(3NN) & -2.72 & -0.57 & -0.85 & -29.08 & 1.4 & 422\\
&x=0.25(NN) & -2.74 & -0.56 & -0.87 & -29.30 & 1.44 & 423\\
\hline
\end{tabular}
\caption{Exchange interaction strengths J$_{1}$ (NN), J$_{2}$ (2NN) and J$_{3}$  (3NN) and effective exchange coupling strength J$_{0}$ in meV, total density of states at Fermi level, D(E$_{F}$) and mean-field Curie Temperature, $T^{mf}_{c}$ for Tl$_{2}$Mn$_{2-x}$Sb$_{x}$O$_{7}$, x=0.0, 0.125 and 0.25. }
\label{J_doping}
\end{table}
Table VI gives the magnetic exchange interaction strengths for all
the three cases and their corresponding T$_{c}$'s and density of
states at E$_{F}$, D(E$_{F}$). For comparison, we also quote the
numbers corresponding to pure Tl$_{2}$Mn$_{2}$O$_{7}$ taken from
ref\cite{dasgupta}. Pristine Tl$_{2}$Mn$_{2}$O$_{7}$ has a very
strong NN FM interaction, a weak 2NN FM interaction and a little
stronger 3NN AFM interaction. With the introduction of doping, all
the interactions became FM and their strengths also increase. On
further increase of doping, the 2nd and 3NN FM interactions become
stronger with a small decrease in the NN FM interaction. It is
worth mentioning here that the results for x=0.25 doping with two
NN Mn atoms substituted, does not change significantly compared to
the result when the two Mn ions substituted were 3NNs, proving
that choice of disordered configuration does not alter the general
trend. Computation of T$_c$, using the values of J$_1$, J$_{2}$
and J$_{3}$, listed in Table I, show rapid increase in T$_c$ on
the initial doping which then attains a kind of saturation on
further doping, considering upto moderate level of doping. This
trend is in good agreement with experimental findings, but the
quantitative values differ.  This happens primarily because
of the clustering effect not included in our calculation. To quote Alonso {\it et. al.}\cite{alonso}, {\it `` For low doping
levels, we cannot disregard some type of magnetic clustering with almost pure Tl$_{2}$Mn$_{2}$O$_{7}$
(T$_c$ = 135K) and Sb rich regions''}. Such effects are bound to reduce the global T$_c$ measured experimentally. Also lattice relaxation effect could be important. Although experimentally measured crystal structure data shows the same space group symmetry as that of the undoped one with little variation in the internal {\it u} parameter, the substitution of Mn ion by a larger cation will probably change the structure locally. Therefore structural relaxations of the doped compounds, which is beyond the scope of our present computational capacity, are needed to resolve the issue completely.

\noindent
\begin{figure}[ht]
\includegraphics[width=8.5cm,keepaspectratio]{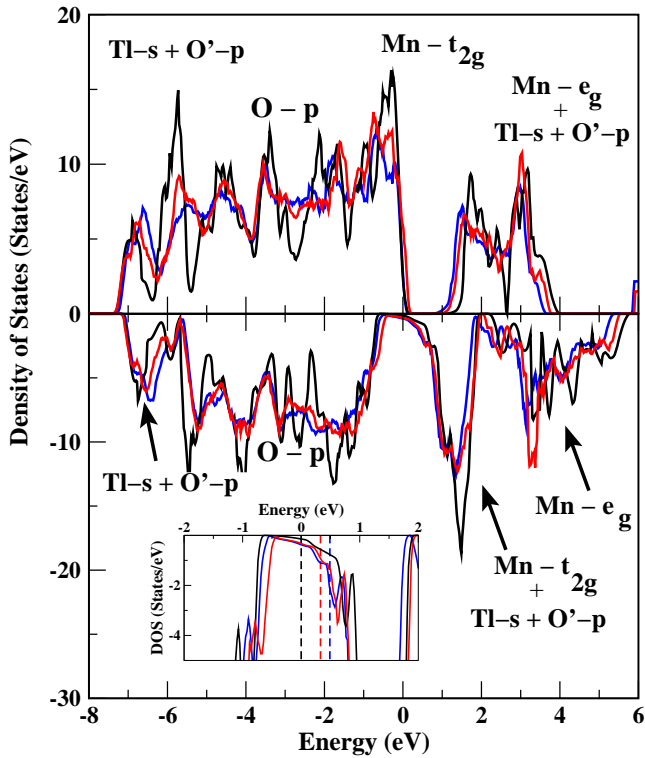}
\caption{(color on-line) Spin-decomposed density of states for x=0.0
(black), 0.125 (red) and 0.25 (blue) doping in
Tl$_{2}$Mn$_{2-x}$Sb$_{x}$O$_{7}$. Zero of the energy is set at the
Fermi level of the undoped Tl$_{2}$Mn$_{2}$O$_{7}$ compound. The upper (lower) panel corresponds to the majority (minority) spin. The negative of density of states has been plotted in the lower panel for clarity. Inset
shows the density of states in minority spin channel at the same
doping levels for the energy region close to Fermi level. The dashed
lines in the inset correspond to the respective Fermi levels.}
\label{dos_dope}
\end{figure}
\par

One needs to examine the electronic structure upon doping to
understand the microscopic origin of the striking effect described
above. In Fig. \ref{dos_dope} we show the density of states
corresponding to Sb-doped Tl$_2$Mn$_2$O$_7$ for doping level of
x=0.125 and 0.25. To appreciate the changes in the electronic
structure upon doping, we also show the density of states for pure
Tl$_{2}$Mn$_{2}$O$_{7}$. Comparing the various density of states, we
observe slight shrinking of the overall band-width due to the small
expansion of the lattice. The t$_{2g}$-e$_g$ splitting at the Mn
site decreases due to expansion of the Mn-O bondlength. Smearing of
the DOS, compared to pure case is also observed due to the effect of
disordering. The Sb ions in the 5+ state have completely filled 4d shell which are essentially core-like while 5s and 5p states remain more or less empty with no contribution to the states close to E$_{F}$. The most important effect is that of electron
doping. Replacement of Mn$^{4+}$ by Sb$^{5+}$ introduces additional
electrons in the system which have no other option but to populate
the highly dispersive Tl-O'-Mn-t$_{2g}$ hybrid state in the minority
spin channel. Because of the highly dispersive nature of this band,
the density of state in the minority spin channel close to E$_F$ rises
fast after the tail part which then attains a kind of small plateau
before attaining highly peaked values corresponding to weakly
dispersive Mn t$_{2g}$ dominated states between 1-2 eV. This is
shown in the inset of Fig.\ref{dos_dope}. With introduction of first
Sb atom, the doped extra electrons while populating the dispersive
minority spin band increases D(E$_F$) substantially. Upon increasing the
number of Sb atoms from one to two, the Fermi level shifts towards
right to accommodate more electrons, but since it has almost crossed
the highly dispersive region and falls within the plateau-like
structure, D(E$_F$) does not change significantly. This is evident
by quantitative estimates of D(E$_F$) shown in Table I. In
Fig.\ref{dos_dope}, we have shown only the case of x=0.25 (NN), but
the DOS corresponding to x=0.25 (3NN) is not very different and
D(E$_F$) changes only marginally. Following the perturbative
treatment in terms of the transfer integral \cite{terakura}, $t$,
between the magnetic and nonmagnetic site, the exchange coupling, J,
within the proposed kinetic energy driven scheme is given by,
$t^{4}D(E_F)/\Delta^{2}$, where $\Delta$ is the average energy
separation between the magnetic and nonmagnetic level and D(E$_F$)
is the density of states at Fermi level introduced in the above. The
real space representation of NMTO-downfolded Hamiltonian keeping
Tl-6s, Mn -3d and O-2p states active and downfolding the rest shows
that the Tl-Mn hopping integral, t's and their energy level
separation changes only slightly for 0.3 $\%$ and 0.9$\%$ expansion
of the lattice upon x=0.125 and 0.25 doping. The major changes come
from the sharp increase in D(E$_F$) which explains the enhancement
of J's and T$_c$'s in turn.

\vskip 1.0cm
\noindent
\section{Conclusions}
\vskip 1.0cm We presented a detailed study of Tl$_2$Mn$_2$O$_7$ under pressure and Sb-doped
Tl$_2$Mn$_2$O$_7$ using
first-principles electronic structure calculations. The analysis
of the computed electronic structure shows that the
counter-intuitive experimental observation of suppression of ferromagnetic T$_c$ upon application of pressure and its enhancement  upon moderate amount of doping by Sb in Mn sublattice is naturally explained in terms of the hybridization induced mechanism proposed earlier\cite{dasgupta}. Application of pressure reduces the hybridization between localized Mn t$_{2g}$ level and the
delocalized Tl-O$^{'}$ effective state due to the shifting of the
position of the Tl-O$^{'}$ effective level, which weakens the
driving mechanism of ferromagnetism and thereby reduces the
strength of exchange coupling and the T$_c$.  
\par
In case of doping,on the other hand, the
enhancement of T$_c$ is induced by the charge-carrier doping of
the system by Sb. This enhances the value of density of states at
the Fermi energy significantly, thereby enhancing the exchange
coupling and the T$_c$. Our conclusions were
substantiated by the explicit calculation of the J's and the
T$_c$'s.
\par
In case of Tl$_2$Mn$_2$O$_7$ under pressure , we resolved the
issue related to the change in Mn-O-Mn angle. We carried out
structural relaxation by means of accurate pseudopotential based
total energy calculations. Our theoretical results show, as
opposed to the claim by Velasco {\it et. al.} \cite{valesco67},
the Mn-O-Mn angle decreases in agreement with the original
suggestion by Sushko {\it et. al.} \cite{sushko}. However the
decrease is only marginal and for all practical purposes, it may
be assumed to remain same as that of Tl$_2$Mn$_2$O$_7$ in ambient
condition. This rules out the use of angle variation argument to
explain the reduction of T$_c$.

\vskip 1.0cm
\noindent
\centerline{\bf Acknowledgements}
\vskip 1.0cm
The research was funded by DST project SR/S2/CMP-42/2003. We thank MPG-partnergroup program for the collaboration.



\end{document}